\documentclass[twocolumn,showpacs,preprintnumbers,aps,amsmath,a4paper,amssymb,amsfont,xcolor=dvipsnames]{revtex4-1}
\pdfoutput=1

\usepackage[pdftex]{graphicx}
\usepackage{graphicx}% Include figure files
\usepackage{dcolumn}% Align table columns on decimal point
\usepackage{bm}% bold math
\usepackage{float}
\usepackage[pdftex,colorlinks=true, citecolor = blue,urlcolor=black,linkcolor=black]{hyperref} %online version
\usepackage{siunitx}
\usepackage{braket}
\usepackage{xcolor}
\DeclareSIUnit\micron{\micro\metre}
\DeclareSIUnit\mrad{\milli\rad}
\begin{document}
\title{Coupled high-finesse optical Fabry-Perot microcavities}	

\author{Steffen Gohlke, Thorsten Langerfeld, Andrea Bergschneider, Michael K{\"o}hl}
\affiliation{Physikalisches Institut, University of Bonn, Wegelerstrasse 8, 53115 Bonn, Germany}

%\date{\today}

\begin{abstract}
Optical fiber Fabry-Perot cavities have been a development facilitating the efficient integration of high-finesse cavities into fiber-optic assemblies. In this work, we demonstrate coupling of two high-finesse fiber cavities by direct photon tunneling between them. We detect the coupled mode spectra and demonstrated the variability of coupling strength and dissipation rates for different transverse modes. Moreover, we observe very narrow spectral features resulting from a dynamical generalizaton of electromagnetically-induced transparency showing that even dissipative systems without metastable states can feature long-lived coherences.

\end{abstract}

\maketitle

%\section{Introduction}
Coupled dissipative quantum systems are an important subject in fundamental science and quantum technologies \cite{Breuer2002}. From the fundamental science perspective, they provide a means to study the effects of open quantum systems and how quantum systems interact with reservoirs. In quantum technologies, coupling dissipative quantum systems is ubiquitous in all kind of interface devices, quantum computers, optomechanics, and hybrid quantum systems. Even though the variety of implementations of coupled dissipative quantum systems is large, there are a few underlying conceptual questions shared between the platforms. Firstly, how to control the dissipation and what effect different levels of dissipation have onto the quantum states. Secondly, how dissipation can be utilized to control and, possibly, enhance quantum coherence \cite{Muller2012}.

\begin{figure}
        \centering
        \includegraphics[width=\columnwidth]{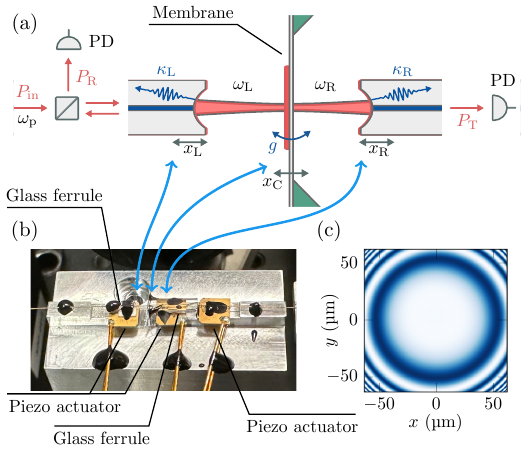}
        \caption{Coupled-cavity setup. (a) Schematic of two coupled Fabry-Perot cavities. (b) Photograph of the experimental setup. The two optical fibers and the center mirror can be moved using shear piezo actuators. (c) Whitelight interference at $\SI{532}{nm}$ of the coated SiN membrane. We extract a symmetric curvature of radius $r_C=\SI{8}{mm}$.}
        \label{fig:fig1}
\end{figure}

A promising platform to investigate the phenomena discussed above are coupled optical cavities, i.e., optical resonators between which photons can coherently tunnel. Coupled cavities \cite{Marcuse1985,Smith2004,Naweed2005,Totsuka2007,Li2010,Stambaugh2015,Liu2017} could act as a model system for open-system control in which one cavity acts as the system and the second cavity is a dissipation channel with tuneable coupling to the system and tuneable dissipation rate. Such a system is described by two resonators with tunable frequencies $\omega_L$ and $\omega_R$ coupled by a coherent coupling rate $g$ via the transmission through the center mirror (see Fig.\,\ref{fig:fig1}\,a). Dissipation from the system happens in the form of photon loss for each of the two resonators with rates $\kappa_L$ and $\kappa_R$ due to finite cavity mirror reflection and scattering or absorption inside the respective resonator which leads to the respective cavity linewidths. Coupled resonators have been experimentally realized on a number of platforms \cite{Liu2017}, among them coupled ring resonators \cite{Naweed2005,Xu2006}, and terahertz metamaterials \cite{Yahiaoui2018}.

In this work, we realize and study two coupled Fabry-Perot resonators based on the optical fiber cavity technology (see Fig.\,\ref{fig:fig1}\,b). To this end, we have prepared an arrangement of two optical fibers, which have been micromachined to have concave surfaces \cite{Hunger2010, Pfeifer2022} and optically coated with a high-reflection coating (T= 100 ppm), centered at $\SI{780}{nm}$ wavelength. The two fiber end facets are $D=\SI{300(10)}{\micro \meter}$ separated from each other. In between the fiber end facets we insert a coupling mirror, which divides the mode volume into two, approximately equally long resonators. The coupling is facilitated by the evanescent wave penetrating through the coupling mirror (see Fig.\,\ref{fig:fig1}\,a). In contrast to previous work with microcavities \cite{Flowers2012}, we utilize a highly reflective mirror separating the cavities, such that the coupling strength between the resonators is on par with the input and output coupling of the resonators, respectively. This coupler between the cavities composes of a free-standing commercial silicon nitride (SiN) membrane of thickness $D_M=\SI{500}{nm}$  with a highly reflective optical coating resulting in a total intensity transmission of $T= \SI{447(52)}{ppm}$, improving significantly over previous work even with macroscopic cavities \cite{Stambaugh2015}. Hence, we expect a coupling parameter of $g\approx 2\pi\times \SI{3.38}{GHz}$. 

An important criterion when designing the coupled-cavity setup is the stability of the cavity modes. Fulfilling the requirement %that both modes should be stable 
becomes more challenging with increasing finesse of the cavities since the photons take more round trips inside the resonator and small imperfections add up. In particular, the alignment and the flatness of the coupling mirror play decisive roles in making both cavities simultaneously stable. We have measured the radius of curvature of the left and the right fiber mirror and the coupler in a white-light interferometer (see Figure \ref{fig:fig1}\,c) and find $r_L=\SI{440}{\micro\meter}$, $r_R=\SI{450}{\micro\meter}$, and $r_C=\SI{8}{mm}$. The flatness of the coupling mirror is strongly influenced by the internal stress of the applied optical coating. By means of a circular mask with $\SI{200}{\micro m}$ diameter, the optical coating was applied only to a circular area, leading to a deviation of only 12\% between the curvatures in vertical and horizontal direction. Attempts to deliberately create internal stress in the SiN membrane during its manufacturing, which would then be compensated when applying the optical coating, have not led to significantly improved results. The alignment between cavity and membrane is achieved by the following procedure: we pre-align the main fiber cavity using two glass ferrules to guide and support the fibers. Then, one fiber is retracted and the membrane is inserted in between the ferrules. We align the membrane using a five-axis translation stage relative to one fiber end facet to form a stable cavity. Subsequently, the retracted fiber is pushed forward again and  we check for stable cavity with the membrane. Finally, the membrane is glued to one of the glass ferrules (see Fig.\,\ref{fig:fig1}\,b).

We characterize our coupled-cavity setup by laser spectroscopy both in reflection and transmission. To this end, we make use of the tunability of the laser frequency and the tunability of both resonator lengths. This allows us to control the detunings $\Delta_p=\omega_p-\omega_R$ and $\Delta_L=\omega_L-\omega_R$ independently, where $\omega_p$ is the probe laser frequency (see Fig.\,\ref{fig:fig1}\,a). We calibrate the linear dependence between cavity length and resonance frequency using side-band modulation. One important aspect to consider is that the nature of the fiber cavity, which comprises of single-mode optical fibers at each end, allows only for an efficient coupling of the TEM$_{00}$ mode of the fibers. Higher-order transverse modes are severely reduced in %intensity 
signal strength at the detector.

\begin{figure}
        \centering
        \includegraphics[width=\columnwidth]{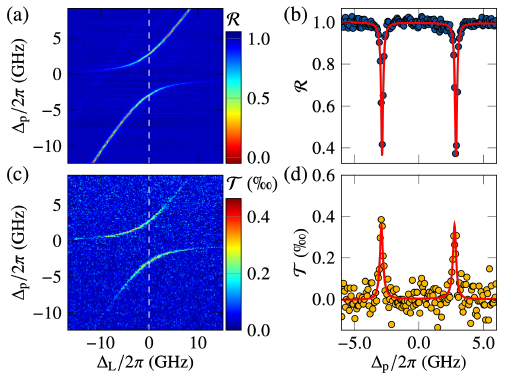}
        \caption{(a), (c) Reflection and transmission spectra of the coupled-cavity system for fixed right cavity length showing an avoided crossing. (b), (d) Reflection and transmission spectrum at $\Delta_L=0$. We observe a mode splitting of $g=2\pi \times\SI{2.859(2)}{GHz}$ at a linewidth of $2\pi\times\SI{189\pm6}{MHz}$.
        }
        \label{fig:fig2}
\end{figure}

We probe the reflection spectrum by pumping the left cavity with $\omega_p$ and detecting the reflected signal. The spectral weight of the signal amplitude $R$ is proportional to the field intensity in the left cavity. Figure \ref{fig:fig2}\,a shows the reflection of the coupled-cavity system as function of the probe laser frequency and the left cavity resonance $\omega_L$, while the right cavity resonance frequency $\omega_R$ remains fixed. When the probe laser is resonant with $\omega_L$ and the right cavity is far detuned, a single resonance is observed with a linewidth caused mainly by cavity losses $\kappa_L$. When the left cavity mode becomes resonant with the right cavity mode, the optical coupling through the center mirror results in an additional phase shift of the field, which leads to the avoided crossing. The mode splitting depends on the field transmission through the center mirror and on the overlap of the spatial modes in the two cavities. At $\omega_L=\omega_R$, the combined cavity modes correspond to the symmetric and anti-symmetric superpositions of the left and right cavity modes and the reflection signal ideally reduces to half of the initial height. We observe a reduction to about 0.4 which points towards non-equal losses $\kappa_L$ and $\kappa_R$. In addition to the two main branches, Figure\,\ref{fig:fig2}\,a  also shows signatures of weaker modes, which we attribute to higher-order transverse modes of different longitudinal modes, see below. 

In Figure \ref{fig:fig2}\,b, we show the transmission spectrum for the same configuration. The transmitted signal is proportional to the intensity in the right cavity and therefore shows the highest signal around zero detuning of the two cavities. This is expected as the right cavity field can be suppressed both by non-resonant coupling from left to right cavity as well as by non-resonant incoupling into the left cavity.

By fitting the spectra with the solutions of a numerical model \cite{Jayich2008}, we extract the normal-mode splitting $2g$ with $g=2\pi \times\SI{2.859(2)}{GHz}$ and a linewidth  $\SI{189\pm 6}{MHz}$. Both cavity modes are TEM$_{00}$ modes as evidenced by their strong coupling to the output fibers. The measured coupling strength is within 20\% of our theoretical expectation, however, the cavity losses are larger by a factor of approximately 2 as compared to the specifications of the optical coating. There are a few effects that contribute to higher losses. We assume equal mirror transmission of the fiber mirrors as they have been produced in the same fabrication process. However, the right cavity contains the SiN-membrane which introduces absorption losses \cite{Jayich2008, Stambaugh2014}. Since the measured reflection spectra allow for an analysis of both linewidths $\kappa_L$ and $\kappa_R$ independently, we can confirm that the cavity containing the SiN substrate has higher losses by approximately $\SI{990\pm176}{ppm}$. This value corresponds to an imaginary part of the refractive index of $2.6(4)\times 10^{-4}$. Additionally, non-perfect alignment of the two cavities and the membrane can lead to different clipping losses for the sub-cavities. Finally, we have observed a gradual reduction of the cavity finesse over time while operating the coupled cavities in air, possibly from contamination.

At the avoided crossing, both coupled cavity modes have equal height and linewidth and, hence, the symmetric and anti-symmetric superposition of left and right modes have equal losses. This agrees with transfer matrix calculations revealing that the field amplitude inside the SiN-membrane is the same for the symmetric and anti-symmetric superposition. 

\begin{figure}
        \centering
        \includegraphics[width=\columnwidth]{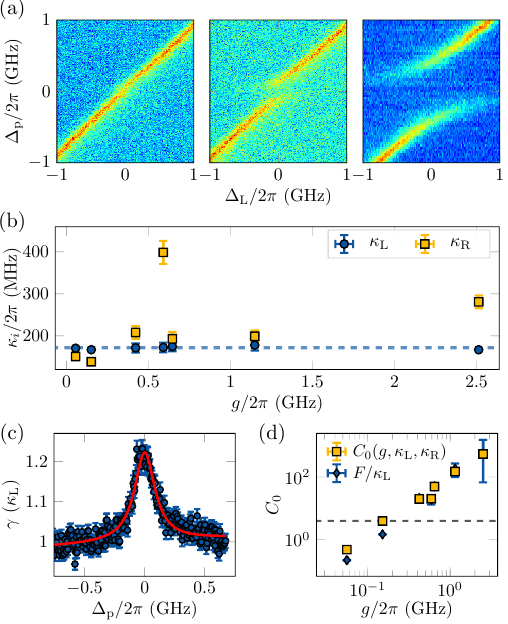}
        \caption{Characterization of the coupled cavities for various coupling strengths. 
        (a) Reflection spectra for higher-order transversal cavity modes in the right cavity result in reduced mode splittings, increasing from left to right. 
        (b) Cavity losses $\kappa_L$ and $\kappa_R$ vs. fitted cavity-coupling $g$. While the losses of the left cavity are constant $\kappa_L=2 \pi\times\SI{172(10)}{MHz}$ over all coupling strengths, varying loss rates in the second cavity are related to the coupling to different transversal modes. 
        (c) Fitted Lorentzian linewidths $\gamma$ from horizontal spectral profiles vs. $\Delta_p$ follow the resonance shape of the right cavity and allow for the independent extraction of $\kappa_R$ from its width and the Purcell factor $4g^2/(\kappa_R)$ from the maximum amplitude.
        (d) Measured cooperativity $C_0=4g^2/(\kappa_L \kappa_R)$ for different cavity couplings reaching over almost three orders of magnitude from weak to strong coupling. The cooperativity extracted from the fitted parameters $\kappa_L$, $\kappa_R$ and $g$ and from the fitted amplitude of the horizontal linewidth are in agreement.}
        \label{fig:fig3}
\end{figure}

The selection of different transverse modes in each cavity provides us with a tuning capability for both the coupling strength $g$ and the losses $\kappa_i$ and therefore gives access to coupling-dominated and loss-dominated regimes. Figure \ref{fig:fig3}\,a shows spectra of higher-order transverse modes in the right cavity that are coupling to the same higher-order transverse mode in the left cavity. The normal-mode splitting for this case is reduced as the spatial mode overlap reduces significantly when selecting different transverse modes in either cavity.

We  analyze the two-dimensional spectra containing higher modes by fitting the vertical profiles to the numerical model in order to extract the coupling strength $g$ and the dissipation rates $\kappa_i$. Far away from the avoided crossings, where photon coupling to the right cavity is suppressed, the observed lineshape is governed by the photon losses of the left cavity $\kappa_L$ only. Note that no photons are lost into the right cavity in this situation. We observe similar linewidths for all measured reflection spectra coupling to this mode resulting in an average value of $\kappa_L= 2 \pi\times \SI{172 \pm 10}{MHz}$ (see Fig.\,\ref{fig:fig3}\,b, blue dashed line). Approaching the zero crossing, more and more field is coherently coupled from the left into the right cavity mode. As a consequence, the losses of the system cross over along with the detuning $\Delta_L$ from purely $\kappa_L$ to $\kappa_R$ and thus the linewidths of the two resonances around the avoided crossing vary along $\Delta_L$ if $\kappa_L$ and $\kappa_R$ are not equal. We fit the spectral profiles with our numerical model to determine $\Delta_L=0$ and obtain $g$ and $\kappa_R$. We measure a number spectra with various couplings strength and thereby observe a strong variation of extracted $\kappa_R$ depending on which transverse modes we couple to (see Fig.\,\ref{fig:fig3}\,b, yellow data points), which we attribute to varying clipping losses for the different higher-order transversal modes in the right cavity. 

We confirm this by directly extracting the losses $\kappa_R$ from the reflection spectra in our coupled-cavity system. For this we analyse horizontal profiles of the reflection spectra which corresponds to choosing a given detuning between the fixed right cavity resonance and the pump laser energy $\Delta_p$ and then scanning the left cavity length.  If $\omega_p$ is far detuned from $\omega_R$, i.\,e. $|\Delta_p|\gg0$, we observe a resonance at $\omega_L$ with a linewidth $\gamma$ identical to $\kappa_L$, as only losses through the left mirror play a role. If $\Delta_p\approx0$, photons inside the left cavity, can also enter the right cavity. This opens an additional dissipation channel for photons, resulting in an increase of the linewidth $\gamma$. Along $\Delta_p$, this increase in dissipation follows the resonance shape of the right cavity mode, as we effectively scan over its resonance and, hence, we can extract $\kappa_R$ from this resonance width (see Fig\,\ref{fig:fig3}\,c). Interestingly, the loss rate $\gamma$ gets enhanced by $F=4g^2/\kappa_R$ compared to $\kappa_L$ at $\Delta_p=\Delta_L=0$, where $F$ is equivalent to the so-called Purcell factor \cite{Purcell1946} for the loss-dominated regime. For the coupling-dominated regime, the shift of the resonance and its decrease in signal strength around $\omega_p=\omega_R$ allows for the extraction of the linewidth only up to a certain point and the maximum amplitude extracted via a Lorentzian fit. We show the quantitative result in terms of the cooperativity $C_0=4 g^2/(\kappa_L \kappa_R)$, which is related to the Purcell factor by $C_0=F/\kappa_L$. {Fig.\,\ref{fig:fig3}\,d shows the cooperativities realized in our system. Hereby, the extraction of $C_0$ from the increase of the linewidth $\gamma$ from the horizontal profiles of the spectra and from the fitted system parameters $g$, $\kappa_L$ and $\kappa_R$ agree well with each other. The results demonstrate that our system allowed for a variation of the cooperativity over almost three orders of magnitude from the dissipation-dominated to the coupling-dominated regime.

When two oscillator modes couple with each other and to a coherent common driving field, interference effects can significantly alter the amplitude and phase of the field and cause a non-linear response. In particular, the spectral response can exhibit features much narrower than the response of the uncoupled systems. In the case of light-matter interactions, this phenomenon is referred to as electromagnetically-induced transparency and for coupled cavity modes, it has been termed coupled-resonator induced transparency (CRIT) \cite{Smith2004,Naweed2005,Liu2017}. EIT and related phenomena have been used to control the dispersion and absorption  of light \cite{Totsuka2007,Fleischhauer2005}.

The standard picture of the EIT/CRIT phenomenon is as follows: there are two dissipative systems with decay rates $\kappa_R$ and $\kappa_L$, respectively, and a coherent coupling strength $g$ between them. If the decay rates are sufficiently different from each other, say $\kappa_L \gg \kappa_R$, and the coupling strength $g$ is significantly less than the faster decay rate $\kappa_L$, spectroscopy of the coupled system will reveal a narrow spectral feature of a width $ 4 g^2/\kappa_L \ll \kappa_L$. For stationary case, however, none of the above conditions for the observability of EIT are fulfilled in our system, since we have $\kappa_R \simeq \kappa_L$ and $g \gg \kappa_L, \kappa_R$.

We demonstrate here a generalization of the EIT phenomenon, which we create and control by scanning the resonance frequency of both cavities simultaneously across each other's resonance. To this end, we realize an asymmetric scan of the two cavity lengths in which the cavity resonance frequencies are scanned into opposite directions and with different rates while the probe laser frequency is  $\omega_p$. In Figure \ref{fig:fig4}\,a we show the reflection spectra of the right cavity for different probe laser detunings. 
One observes a narrow spectral feature  for which the cavity reflection is  increased or decreased depending on the combination of detunings. This feature results from the dynamical effect of sweeping the two cavities across the mutual resonance into opposite directions, see Figure 4\,b. We compare our results to numerical simulations of the coupled resonator fields in which we assume the cavity fields to be in equilibrium since the pump and decay rates of the cavities are much faster than the rate of change of the relative detunings. Our results therefore show that one can realize dynamically very narrow spectral transparencies in the coupled-cavity arrangement despite not fulfilling the requirements for stationary EIT/CRIT. This is significantly different from standard EIT physics, in which at least one long-lived state is required in order to observe the narrow spectral features and could be used, for example, for optical switching or optical mode selection.

\begin{figure}
        \centering
        \includegraphics[width=\columnwidth]{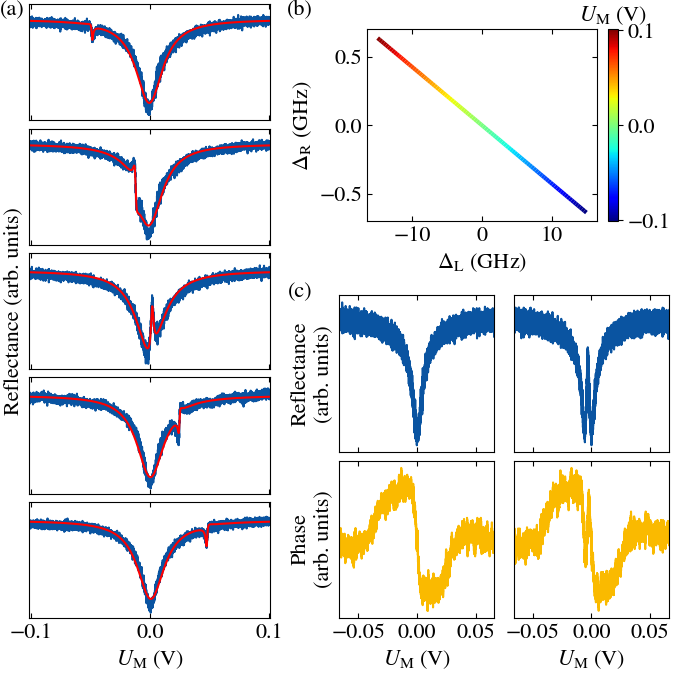}
        \caption{Dynamical coupled-resonator induced transparency. (a) Coupled-cavity spectrum for different detunings $\Delta_L$ and piezo scan of the membrane $U_M$. (b) Detuning of left and right cavity resulting from the membrane modulation with $U_M$. The friction of the mounting ferrule causes the reduced resonance scan of the right cavity $\Delta_R$.  (c) Reflection and its Pound-Drever-Hall signal for off-resonant (left) and on-resonant (right) coupled-cavity system. The latter is proportional to the phase. (d) Corresponding phase shift of the reflectance signal.}
        \label{fig:fig4}
\end{figure}

In Figure \ref{fig:fig4}\,c, we show a detailed scan of both the reflectance and the phase of the reflected field with and without the EIT phenomenon. The two left plots show the results without the EIT effect and one observes the Lorentzian lineshape in the reflectance and the dispersive signal for the variation of the phase. The two right plots show the signal in presence of the EIT phenomenon and one observes the narrow spectral feature appearing both in reflectance and phase measurements.

In summary, we have realized and characterized an arrangement of two coupled high-finesse Fabry-Perot optical cavities. Spectroscopy of the coupled cavities reveals the avoided crossing of the coupled cavity modes. Given by the high reflectivity of the cavity and coupling mirrors, we find a coupling strength of only a couple of GHz, which, in the future, could make the real-time observation of photon tunnelling between the two cavities possible. Further, the excellent transverse access to the cavity modes in our realization could enable the introduction of optical non-linearities, such as single atoms, into the cavities towards the quantum simulation of lattice models using interacting photons \cite{Hartmann2006,Hartmann2016}.

%\section{Acknowledgements}
This work has been supported by BMBF (project QR.X), Cluster of Excellence Matter and Light for Quantum Computing (ML4Q) EXC 2004/1–390534769, and Deutsche Forschungsgemeinschaft (SFB/TR 185, project A2).

\bibliographystyle{apsrev4-1}

\end{document}